\documentclass{WileyMSP-template}

\usepackage{color}
\begin{document}
\linespread{2.0}


\title{Highly Efficient Room-Temperature Nonvolatile Magnetic Switching by Current in Fe$_{3}$GaTe$_{2}$ Thin Flakes}
\maketitle

\author{Shaohua Yan,$^{1,2,\dag}$ Shangjie Tian,$^{3,1,2,\dag}$ Yang Fu,$^{1,2}$ Fanyu Meng,$^{1,2}$ Zhiteng Li,$^{4}$ Hechang Lei,$^{1,2,*}$, Shouguo Wang,$^{3,*}$, and Xiao Zhang$^{4,*}$}

\begin{affiliations}
S. H. Yan, Dr. S. J. Tian, Y. Fu, F. Y. Meng, Prof. H. C. Lei\\
Department of Physics and Beijing Key Laboratory of Optoelectronic Functional Materials \& MicroNano Devices, Renmin University of China, Beijing 100872, China\\
Email Address: hlei@ruc.edu.cn

S. H. Yan, Dr. S. J. Tian, Y. Fu, F. Y. Meng, Prof. H. C. Lei\\
Key Laboratory of Quantum State Construction and Manipulation (Ministry of Education), Renmin University of China, Beijing, 100872, China\\
Email Address: hlei@ruc.edu.cn
 
Dr. S. J. Tian, Prof. S. G. Wang\\
Anhui Key Laboratory of Magnetic Functional Materials and Devices, School of Materials Science and Engineering, Anhui University, Hefei 230601, China\\
Email Address: sgwang@ahu.edu.cn

Z. T. Li, Prof. X. Zhang\\
State Key Laboratory of Information Photonics and Optical Communications $\&$ School of Science, Beijing University of Posts and Telecommunications, Beijing 100876, China\\
Email Address: zhangxiaobupt@bupt.edu.cn
\end{affiliations}


\newpage
\begin{abstract}
\bf{Effectively tuning magnetic state by using current is essential for novel spintronic devices. Magnetic van der Waals (vdW) materials have shown superior properties for the applications of magnetic information storage based on the efficient spin torque effect. However, for most of known vdW ferromagnets, the ferromagnetic transition temperatures lower than room temperature strongly impede their applications and the room-temperature vdW spintronic device with low energy consumption is still a long-sought goal.
Here, we realize the highly efficient room-temperature nonvolatile magnetic switching by current in a single-material device based on vdW ferromagnet Fe$_{3}$GaTe$_{2}$. Moreover, the switching current density and power dissipation are  about 300 and 60000 times smaller than conventional spin-orbit-torque devices of magnet/heavy-metal heterostructures.
These findings make an important progress on the applications of magnetic vdW materials in the fields of spintronics and magnetic information storage.}
\end{abstract}
~\\
Two-dimensional magnetic materials are new members of the two-dimensional material family. They still maintain long-range magnetic order at the thickness of a single unit cell and are easily controlled by external fields \cite{Gong1,Bevin1,Deng,ZhangGJ,TianSJ,LeeJU,LeeK}. This provides an ideal platform for the study of magnetism and other novel physical effects under two-dimensional limits. Moreover, owing to their intrinsic two-dimensional (2D) long-range magnetism and the easiness of combination with other van der Waals (vdW) materials, the vdW 2D magnets also exhibit significant advantages in the design and preparation of multi-functional vdW devices with reduced size of device, especially spintronic devices \cite{Geim,Novoselov,LiuY}. Thus they have been attracted intensive attention.
Among the known vdW magnetic materials, the Fe-based systems, such as Fe$_{x}$GeTe$_{2}$ ($x$ = 3, 4, 5) and Fe$_{3}$GaTe$_{2}$, exhibit superior properties like high magnetic transition temperature ($T_{c}$) up to about 380 K and large coercivity of approximately several kOe at low temperature with the $c$ axis magnetic anisotropy \cite{ZhangGJ,Deiseroth,May,SeoJ}. More importantly, such features of hard magnetism with high $T_{c}$ persist in the atomically thin samples and can be tuned via various methods, such as doping, pressure, gating etc. \cite{Deng,ZhangGJ,LiZ}.
~\\
Effective control of magnetic state by using current is at the core of spintronics field \cite{Myers,WangQY}. The adjustable coercivity by current makes it possible to simultaneously achieve high efficiency and nonvolatile characteristics in devices.
For an information writing process, the coercive field ($H_{c}$) need be reduced to relatively low value and the information can be written with low energy consumption. On the other hand, once the information has been written, the $H_{c}$ should be big enough to reduce the information loss caused by fluctuations. 
Previous studies demonstrate that the nonvolatile and energy-efficient magnetization switching by current can be achieved in single vdW ferromagnet Fe$_{3}$GeTe$_{2}$ \cite{ZhangKX,ZhangKX2}.
But the $T_{c}$ of Fe$_{3}$GeTe$_{2}$ lower than room temperature limits its application.
In contrast, the $c$-axial vdW hard ferromagnet Fe$_{3}$GaTe$_{2}$ has a much higher $T_{c}$ than 300 K. 
Moreover, based on Fe$_{3}$GaTe$_{2}$ thin flakes, some room-temperature magnetic tunneling junction and spin-orbit torque (SOT) devices have been reported \cite{ZhuWK, YinHF, JinW, LiWH}.

~\\
Motivated by the studies of current-controlled magnetic switching in Fe$_{3}$GeTe$_{2}$ and the superior properties of Fe$_{3}$GaTe$_{2}$, in this work, we study the evolution of coercive field $H_{c}$ with the $ab$-plane current for Fe$_{3}$GaTe$_{2}$ thin flakes. The $H_{c}$ can be reduced remarkably using the relatively low current density and importantly the current-controlled $H_{c}$ still exists at room temperature. Furthermore, the nonvolatile magnetic switching by current  is realized at room-temperature with high efficiency. 
Switching current density and power dissipation at 300 K are  about 300 and 60000 times smaller than conventional \linebreak
ferromagnet/heavy-metal systems, 
Such a high-efficiency room-temperature magnetic memory device could be a key component in the area of all-vdW spintronics.

~\\
The Fe$_{3}$GaTe$_{2}$ thin flakes were exfoliated from high-quality bulk crystals grown by self-flux method. As is shown in Fig. 1a, Fe$_{3}$GaTe$_{2}$ has a layered structure with a hexagonal structure (space group $P6_{3}/mmc$) \cite{ZhangGJ}, isostructural to Fe$_{3}$GeTe$_{2}$, and the fitted lattice parameters $a$ and $c$ from the single-crystal X-ray diffraction (XRD) pattern is 0.409(2) nm and  1.607(2) nm, respectively, which are consistent with previous results \cite{ZhangGJ}. In the Fe-Ga-Te layer, the Fe-Ga slab is sandwiched by the Te atoms and there is a vdW gap between  two adjacent Fe-Ga-Te slabs, thus it can be exfoliated easily. 
Fig. 1b shows the XRD pattern of a Fe$_{3}$GaTe$_{2}$ single crystal and only sharp (00$l$) diffraction peaks are observable, indicating that the crystal surface is normal to the $c$-axis with the plate-shaped surface parallel to the $ab$-plane. 
Fig. 1c displays a typical device with a Hall bar geometry. The device is made by attaching the Fe$_{3}$GaTe$_{2}$  thin flake on a pre-prepared Ti/Au electrode and covered by h-BN for protection from H$_{2}$O and O$_{2}$. 
The thickness $t$ of this sample determined by the atomic force microscopy (AFM) (Fig. 1d) is around 12 nm. 
As shown in Fig. 1e, Fe$_{3}$GaTe$_{2}$ with $t$ = 12 nm shows a metallic behavior in the range of 18 K - 320 K, and there is Kondo-like minimum at $\sim$ 18 K, similar to the results of Fe$_{3}$GaTe$_{2}$ thin flakes in the literature \cite{ZhangGJ}. 
Fig. 1f exhibits the $c$-axial field $H_{z}$ dependence of Hall resistivity $\rho_{yx}(H_{z})$ at 300 K with current $I$ = +0.01 mA and -0.01 mA applied along the $ab$-plane. 
The obvious hysteresis loop in the $\rho_{yx}(H_{z})$ curves clearly indicates that the room-temperature $c$-axial hard ferromagnetism persists in the Fe$_{3}$GaTe$_{2}$  thin flake.  
Moreover, it can be seen that with small $I$ (= $\pm$ 0.01 mA), the $\rho_{yx}(H_{z})$ curves with opposite current directions are almost same shape and both of them have identical $H_{c}$ ($\sim$ 70 Oe).

~\\
Fe$_{3}$GaTe$_{2}$ thin flakes exhibit an obvious behavior of current-dependent $H_{c}$. Taking the device with $t$ = 12 nm (sample 1, S1) as an example (Fig. 2a),  at $T$ = 5 K, the hysteresis loop becomes smaller with the decreased $H_{c}$ when the $I$  increases. The $H_{c}$ is around 2.5 kOe under $I$ = 0.01 mA and decrease to 1.8 kOe when a 9 mA current is applied (Fig. 2b). 
In contrast to the significant changes of $H_{c}$, the saturation Hall resistivity $\rho_{yx}^{s}$ related to the saturation magnetization is almost intact. 
Importantly, as shown in Figs. 2c and 2d, the current-controlled $H_{c}$ still exists at room temperature ($T$ = 300 K). For $I$ = 0.01 mA, the $H_{c}$ is $\sim$ 70 Oe and with increasing current the $H_{c}$ decreases gradually. Finally it becomes almost zero at $I$ = 3 mA. On the other hand, the  $\rho_{yx}^{s}$ is barely changed with current, similar to the results at 5 K. It implies that the ferromagnetic order is not been affected by the $ab$-plane current and thus Fe$_{3}$GaTe$_{2}$ is suitable for spintronics at room temperature.
Fig. 2e displays the evolution of reduced $H_{c}(I)/H_{c}(I = 0.01 \rm mA)$ vs. $I$ at different temperatures.  The curves at the low-temperature region ($T$= 5 - 200 K) exhibit similar behavior while  the $H_{c}$ decreases with increasing current more quickly at 250 K and 300 K.
The larger current tunability of $H_{c}$ at both temperatures can be partially ascribed to the weakened ferromagnetism because of the stronger thermal fluctuations at higher temperature.
To confirm the reproducibility of properties of our devices, we measured several samples with different thicknesses (S1 - S3). They exhibit a similar reduction of $H_{c}$ via $I$ and the $H_{c}$ of thinner sample decreases with $I$ more quickly (Fig. S1 in Supporting Information).
Fig. 2f shows the relationship between reduced  $H_{c}(J)/H_{c}$($J \sim$ 0.01 mA/$\mu$m$^{2}$) and current density at 300 K. It can be seen that all of curves for three devices almost fall on the same line, and the current reduces $H_{c}$ by $\sim$ 50 \% for $\sim$ 5 mA/$\mu$m$^{2}$
when the $H_{c}$ disappears at around $J$ = 12.5 mA/$\mu$m$^{2}$.

~\\
Based on the current-controlled coercivity reduction, we demonstrate that a highly energy-efficient nonvolatile spin memory device can be realized  in Fe$_{3}$GaTe$_{2}$ thin flakes at room temperature, in which magnetic information is written by $I$ and read through the $\rho_{yx}(H_{z})$.
Fig. 3a shows the hysteresis loop of $\rho_{yx}(H_{z})$ under 0.01 mA and 9 mA at 5 K. The negative and positive saturated magnetization states $\pm M_{s}$ are defined as ``0'' and ``1" states.
When sweeping the field from  -5 kOe to 2.2 kOe, the initial state of device is on the ``0" state. Then a large writing current $I$ of 9 mA is applied and because the $H_{c}$ is significantly reduced by $I$, the magnetization of the Fe$_{3}$GaTe$_{2}$ device can be switched by current from ``0" state to ``1" state (blue arrow). Correspondingly, the ``1'' state at -2.2 kOe initially can also be altered to ``0" state by the large $I$ (9 mA) (red arrow). 
Fig. 3b shows the detailed switching process of states through the $\rho_{yx}(H_{z})$ - $\rho_{yx}$(-5 kOe) as a function of $I$ path (0 $\rightarrow$ 9 $\rightarrow$ -9 $\rightarrow$ 9 mA) under various fields from -5 kOe to 5 kOe. For the ``0" state at $H_{z}$ = -5 kOe and the ``1" state at $H_{z}$ = 5 kOe in Fe$_{3}$GaTe$_{2}$ device, they are stable regardless of current sweep (purple and green curves). When fixing the field at 2.2 kOe, the increase of $I$ from 0 to 9 mA leads to the change of magnetization gradually from -$M_{s}$ (``0" state) to +$M_{s}$ (``1" state).
Importantly, after reaching the ``1" state, the magnetization is unchanged even the $I$  sweeps back to - 9 mA and forth 9 mA again. It is clearly indicate that the magnetic information switching from ``0" to ``1" states through  $I$  is robust and nonvolatile (blue curve).
Similarly, the reverse switching from ``1" to ``0" states can also be realized using the same $I$ path with the field at -2.2 kOe (red curve). 
On the other hand, the magnetic information (the value of $\rho_{yx}(H_{z})$) can be read using a small current without disturbing the magnetization states. 
Similar results are also observed when $H_{z}$ = $\pm$1.8 kOe and $\pm$2.0 kOe (Fig. S2 in Supporting Information), indicating that there is a relatively wide window of magnetic field to switch the ``0" and ``1" states by current. 
More importantly, such switching of magnetic states driven by current persists up to room temperature  ($T$ =  300 K) (Figs. 3c and 3d). 
As shown in Fig. 3d, the ``0" state at $H_{z}$ = -1 kOe and the ``1" state at $H_{z}$ = 1 kOe are robust to the switching process $I$ (0 $\rightarrow$ 3 $\rightarrow$ -3 $\rightarrow$ 3 mA) (purple and green curves). 
When fixing the field at $\pm$20 Oe, the increase of $I$ from 0 to 3 mA can switch the magnetization gradually, similar to the case at 5 K.
In addition, the minimum $I$ ($\sim$ 1.2 mA) changing the ``0" and ``1" states at 300 K is much smaller than that at 5 K ($\sim$ 6 mA). Meanwhile, the field where the current can switch magnetic states also becomes very low (below 100 Oe). These behaviors are consistent with the significant reduction of $H_{c}$ with $I$ at 300 K (Fig. 2e).
It is noted that at 300 K the hysteresis loop of $\rho_{yx}(H_{z})$ curve at 0.01 mA is not a rectangular shape  and it disappears at 3 mA, thus the initial and final states at $\pm$20 Oe is slightly different from the  ``0" and ``1" states at $\pm$1 kOe (green arrows in Figs. 3c and 3d). But the switching between the initial and final states at $\pm$20 Oe is still large enough to be discerned. 

~\\
Next, we try to explain the electrical modulation of $H_{c}$ qualitatively based on the SOT effect generated by an $ab$-plane current in Fe$_{3}$GaTe$_{2}$ itself due to its special geometrical structure same as Fe$_{3}$GeTe$_{2}$ \cite{ZhangGJ}. 
To exclude the influence of Joule heating effects, the $R_{xx}(T)$ curves from 280 K to 320 K with $I$ = 0.01 mA and 3 mA have been measured (Fig. S3) and these two curves are almost identical, indicating that the heat effect should be negligible.  So the reduce of coercive field and the magnetic switching can be mainly attributed to SOT effect.
Previous studies on Fe$_{3}$GeTe$_{2}$ indicate that an $ab$-plane current density $\textbf{\textit{J}}$ = ($J_{x}$, $J_{y}$, 0) can induce a SOT $\textbf{\textit{T}}_{\rm SOT}$ = -$|\gamma|\textbf{\textit{M}}\times \textbf{\textit{H}}_{\rm SOT}$ acting on $\textbf{\textit{M}}$ = ($M_{x}$, $M_{y}$, $M_{z}$), where $|\gamma|$ is the gyromagnetic ratio and the $\textbf{\textit{H}}_{\rm SOT}$ is the effective SOT magnetic field \cite{ZhangKX,Johansen}. 
And the symmetry of the crystal determines the dependence of $\textbf{\textit{H}}_{\rm SOT}$ on $\textbf{\textit{m}}$ and $\textbf{\textit{J}}$ and it can be expressed as $\textbf{\textit{H}}_{\rm SOT}$ = $\Gamma_{0}$[($m_{x}J_{x}$ - $m_{y}J_{y})$$\textbf{\textit{e}}_{x}$ - ($m_{y}J_{x}$ + $m_{x}J_{y})$$\textbf{\textit{e}}_{y}$], where $\textbf{\textit{m}}$ = ($m_{x}$, $m_{y}$, $m_{z}$) = $\textbf{\textit{M}}$/$|\textbf{\textit{M}}|$ is the magnetization unit vector and $\Gamma_{0}$ is the strength of the magnetoelectric coupling, determined by the spin-orbital coupling \cite{ZhangKX,Johansen}. 
Moreover, the $\textbf{\textit{H}}_{\rm SOT}$ can be obtained from an effective free energy density $f_{\rm SOT}$ by $\textbf{\textit{H}}_{\rm SOT}$  = -$\partial f_{\rm SOT}$/$\partial \textbf{\textit{M}}$ with $f_{\rm SOT}$ = $M_{s}\Gamma_{0}[J_{y}m_{x}m_{y} - \frac{1}{2}J_{x}(m_{x}^{2} - m_{y}^{2})]$.
Combined with the magnetocrystalline anisotropy energy (MAE) for Ising-type out-of-plane easy-axis ferromagnet $f_{\rm MAE}$ = -$\frac{1}{2}K_{z}M_{z}^{2}/M_{s}$, we have the zero-field effective free energy density $f_{\rm eff}$ = $f_{\rm MAE}$ + $f_{\rm SOT}$. 
If considering a spatially uniform magnetization and use a spherical basis, ($m_{x}$, $m_{y}$, $m_{z}$) = (sin$\theta$cos$\phi$, sin$\theta$sin$\phi$, cos$\theta$) and ($J_{x}$, $J_{y}$, $J_{z}$) = $|J|$(cos$\phi_{J}$, sin$\phi_{J}$, 0), it has $ f_{\rm eff}$ = -$\frac{1}{2}M_{s}$[$K_{z}$cos$^2$$\theta$ + $\Gamma_{0}J$sin$^2$$\theta$cos(2$\phi$ + $\phi_{J}$)] \cite{ZhangKX,Johansen}.
For $J$ = 0, the $f_{\rm eff}$ has a minimum value of -$M_{s}K_{z}$/2 at $\textbf{\textit{m}}$ = $\pm\textbf{\textit{e}}_{z}$ and a maximum value of 0 when $\textbf{\textit{m}}$ lies in the $xy$ plane. 
Therefore, the free energy barrier for switching the magnetization from $\textbf{\textit{m}}$ = -$\textbf{\textit{e}}_{z}$ to $\textbf{\textit{m}}$ = +$\textbf{\textit{e}}_{z}$ is $M_{s}K_{z}$/2. 
In contrast, the existence of $ab$-plane $J$ modifies this free energy barrier to $M_{s}$($K_{z}$ - $|\Gamma_{0}J|$)/2, i.e., the $J$ effectively reduce the barrier height between the local minima of the free energy \cite{ZhangKX}. 
Correspondingly, the coercive field $H_{c}$ is also reduced from $K_{z}$ at $J$ = 0 to $K_{z}$ - $|\Gamma_{0}J|$ at finite $ab$-plane $J$ in a single domain case. 
This linear behavior is clearly observed from the $H_{c}(J)$ curves 
for S1 -  S3 of Fe$_{3}$GaTe$_{2}$ at 300 K (Fig. 4a).
The linear fits of $H_{c}(J)$ curves at the low current density range give $|\Gamma_{0}|$= 7.0(3), 5.7(3), and 8.4(5) Oe/(mA/$\mu$m$^{2}$), respectively.  
These $|\Gamma_{0}|$ values are comparable with Fe$_{3}$GeTe$_{2}$/Pt bilayer system \cite{WangX}, and one order of magnitude larger than those in the conventional heavy metal multiplayer films, such as Pt/Co and Pt/Fe etc \cite{ZhangKX,WangX,Khang,HanJ,Garello,M} (Fig. 4b).
Moreover, although the $|\Gamma_{0}|$ of Fe$_{3}$GaTe$_{2}$ in present work is somewhat smaller than that of Fe$_{3}$GeTe$_{2}$ device \cite{ZhangKX}, the work temperature of Fe$_{3}$GaTe$_{2}$ device at 300 K make it significantly superior to Fe$_{3}$GeTe$_{2}$.
On the other hand, lowering switching current density $J_{\rm SW}$ and power dissipation $J_{\rm SW}^{2}/\sigma$ ($\sigma$ = 1/$\rho$ is conductivity) are important to improve the energy efficiency of memory device. 
These two parameters of three Fe$_{3}$GaTe$_{2}$ device have been calculated and shown in Fig. 4c. 
At 300 K, the $J_{\rm SW}$ and $J_{\rm SW}^{2}/\sigma$ are in the range of 4.7(1) - 7.7(3)$\times$10$^{5}$ A/cm$^{2}$ and 1.5(2) - 2.0(1)$\times$10$^{13}$ W/m$^{3}$, both of which are significantly smaller than most of composite SOT systems.
For example, the $J_{\rm SW}$ and $J_{\rm SW}^{2}/\sigma$ of Fe$_{3}$GaTe$_{2}$ are about 300 and 60000 times smaller than conventional heavy-metal/magnet SOT Pt/Co system \cite{ZhangKX2}.  
In addition, when compared with Fe$_{3}$GeTe$_{2}$, both of them have similarly small $J_{\rm SW}$ but the $J_{\rm SW}^{2}/\sigma$ of Fe$_{3}$GaTe$_{2}$ is one order of magnitude lower than that of  Fe$_{3}$GeTe$_{2}$, highlighting the high energy efficiency of present Fe$_{3}$GaTe$_{2}$ devices.

The existence of such SOT effect can be further verified by the current-driven magnetization switching with the assistance of an $ab$-plane magnetic field $H_{x}$ along the current direction at 300 K (Fig. 5). The magnetization of Fe$_{3}$GaTe$_{2}$ can be switched from one state to the other by sweeping the electric current $I_{x}$. The opposite sweeping direction of $I_{x}$ leads to the opposite switch of magnetization and the switching polarity is also inverse from anticlockwise to clockwise when the $H_{x}$ is changed from 500 Oe to -500 Oe. Such behavior is very similar to those observed in Pt/Fe$_{3}$GeTe$_{2}$ \cite{WangX}, Pt/CoFe/MgO and Ta/CoFe/MgO \cite{Emori}, which have been explained well by the SOT effect. It demonstrates undoubtedly that the present current-induced switch of magnetization is mainly due to the SOT effect. Furthermore, it is noted that the current where the magnetic moments start to flip is relatively small, which is consistent with the small flipping current shown in Fig. 3.
Moreover, as shown in Fig. S4, we has also measured the evolution of $\rho_{yx}$($H_{z}$ = 0 Oe) - $\rho_{yx}$(-1 kOe) of sample S1 as a function of current $I$ path (0 $\rightarrow$ 3 $\rightarrow$ -3 $\rightarrow$ 3 mA)  with initial saturated magnetization states at +$M_{s}$ and -$M_{s}$ at 300 K and $H_{x}$ = 0 Oe. It can be clearly seen that when the $H_{x}$ and $H_{z}$ both are zero, the sweeping of $I$ from 0 to 2 mA can only change the $\rho_{yx}$($H_{z}$ = 0 Oe) -  $\rho_{yx}$(-1 kOe) to the half value of  $\rho_{yx}$(+1 kOe) -  $\rho_{yx}$(-1 kOe) and further sweeping $I$ from +3 mA to -3 mA will not change its value. It indicates that without the assistance of $H_{x}$ or $H_{z}$, the spins due to the SOT effect can only be switched from $c$ axis to $ab$ plane only, in agreement with the theoretical result \cite{Johansen} . 
 

~\\
In summary, the electrical control of coercive field in Fe$_{3}$GaTe$_{2}$ thin flake devices is observed up to room temperature. 
In addition, the room-temperature nonvolatile magnetization switching behavior at very low field ($\sim$ 20 Oe) with a small $ab$-plane current density ($\sim$ 5.0$\times$10$^{5}$ A/cm$^{2}$) and the remarkably low power dissipation ($\sim$ 1.50 $\times$ 10$^{13}$ W m$^{-3}$) is realized.
These behaviors can be partially ascribed to the unique SOT effect in Fe$_{3}$GaTe$_{2}$.
Such capability of high-efficiency nonvolatile magnetization switching by current  at room temperature make Fe$_{3}$GaTe$_{2}$ become a very promising system for spintronic applications using 2D magnetic vdW materials. 

~\\
\textbf{Experimental Section}

~\\
\noindent\textbf{Single crystal growth and structural characterization.} 
Single crystals of Fe$_{3}$GaTe$_{2}$ were grown by self-flux method. Flakes of Fe (99.98 \% purity), Ga (99.99 \% purity) and Te (99.99 \%) in a molar ratio of 1 : 1 : 2  were put into a quartz tube. The tube was evacuated and sealed at 0.01 Pa. The sealed quartz ampoule was heated to 1273 K for 10 hours and are then held there for another one day, then held there for another one day, then the temperature was quickly decrease down to 1153 K witnin 2 h followed by slowly cooled down to 1053 K within 100 h. Finally, the ampoule was taken out from the furnace and decanted with a centrifuge to separate Fe$_{3}$GaTe$_{2}$ single crystals from the flux. In order to avoid degradation, the Fe$_{3}$GaTe$_{2}$ single crystals are stored in an Ar-filled glovebox. The XRD pattern of a Fe$_{3}$GaTe$_{2}$ single crystal was measured using a Bruker D8 Advance X-ray machine with Cu K$\alpha$ ($\lambda$ = 1.5418 \AA) radiation. For the fit of lattice parameters of  Fe$_{3}$GaTe$_{2}$, the single-crystal XRD pattern was measured using a Bruker D8 Quest X-ray machine with Mo K$\alpha$ ($\lambda$ = 0.7107 \AA) radiation. The microscopy images was acquired using a Bruker Edge Dimension atomic force microscope (AFM). 

\noindent\textbf{Device fabrication.} 
Fe$_{3}$GaTe$_{2}$ flakes were cleaved from bulk crystals onto polydimethylsiloxane (PDMS) by mechanical exfoliation and they were examined by an optical microscope to evaluate the thickness roughly. 
Then the atomically smooth flakes with desired thicknesses were transferred to a 285 nm SiO$_{2}$/Si substrate with pre-patterned electrodes and an $h$-BN capping layer was used to cover the sample for protection from H$_{2}$O and O$_{2}$. 
The Ti/Au (10/40 nm) electrodes was fabricated by electron beam lithography and metals were deposited using thermally evaporating method.
After transport measurements, the $h$-BN capping layer was removed and the thickness of sample was determined precisely by AFM. 
The whole fabrication process of device was carried out in an argon glove box with H$_{2}$O and O$_{2}$ contents less than 0.1 ppm to avoid degradation of the samples. 

\noindent\textbf{Electrical transport measurements.} 
Magnetization and electrical transport measurements were performed in a Quantum Design MPMS3 and superconducting magnet system (Cryomagnetics, C-Mag Vari-9). Both longitudinal and Hall electrical resistance were measured using a five-probe method on Fe$_{3}$GaTe$_{2}$ Hall bar device with current flowing in the $ab$ plane. In dc measurements, the bias current was generated by using a current source (Keithley, 6221) and the voltage was measured with a nanovoltmeter (Keithley, 2182A).
The raw Hall resistance was measured by sweeping the field up to $\pm$5 kOe at various temperatures, and the Hall resistance was determined by a standard symmetrization procedure to remove the contribution of magnetoresistance from the raw Hall data due to voltage probe misalignment \cite{Ohno2}.


\medskip
\textbf{Acknowledgments} \par 

This work was supported by  Beijing Natural Science Foundation (Grant No. Z200005), National Key R\&D Program of China (Grants Nos. 2018YFE0202600, 2022YFA1403800 and 2022YFE0109200), National Natural Science Foundation of China (Grants Nos. 12174443 and 12241406), the Outstanding Innovative Talents Cultivation Funded Programs 2022 of Renmin University of China, Beijing National Laboratory for Condensed Matter Physics, and Collaborative Research Project of Laboratory for Materials and Structures, Institute of Innovative Research, Tokyo Institute of Technology.

\textbf{Author contributions} \par
H.C.L., S.G.W. and X.Z. provided strategy and advice for the research; S.H.Y. grew the single crystal; S.H.Y. prepared the devices and performed the measurements of physical properties with the assistance of S.J.T., Y.F., Z.T.L. and F.Y.M.; S.H.Y performed the fundamental data analysis with the assistance of S.G.W., X.Z. and H.C.L.; S.H.Y., H.C.L. and X.Z. wrote the manuscript based on discussion with all the authors.
\medskip

%

\begin{figure}
\centerline{\includegraphics[width=0.8\linewidth]{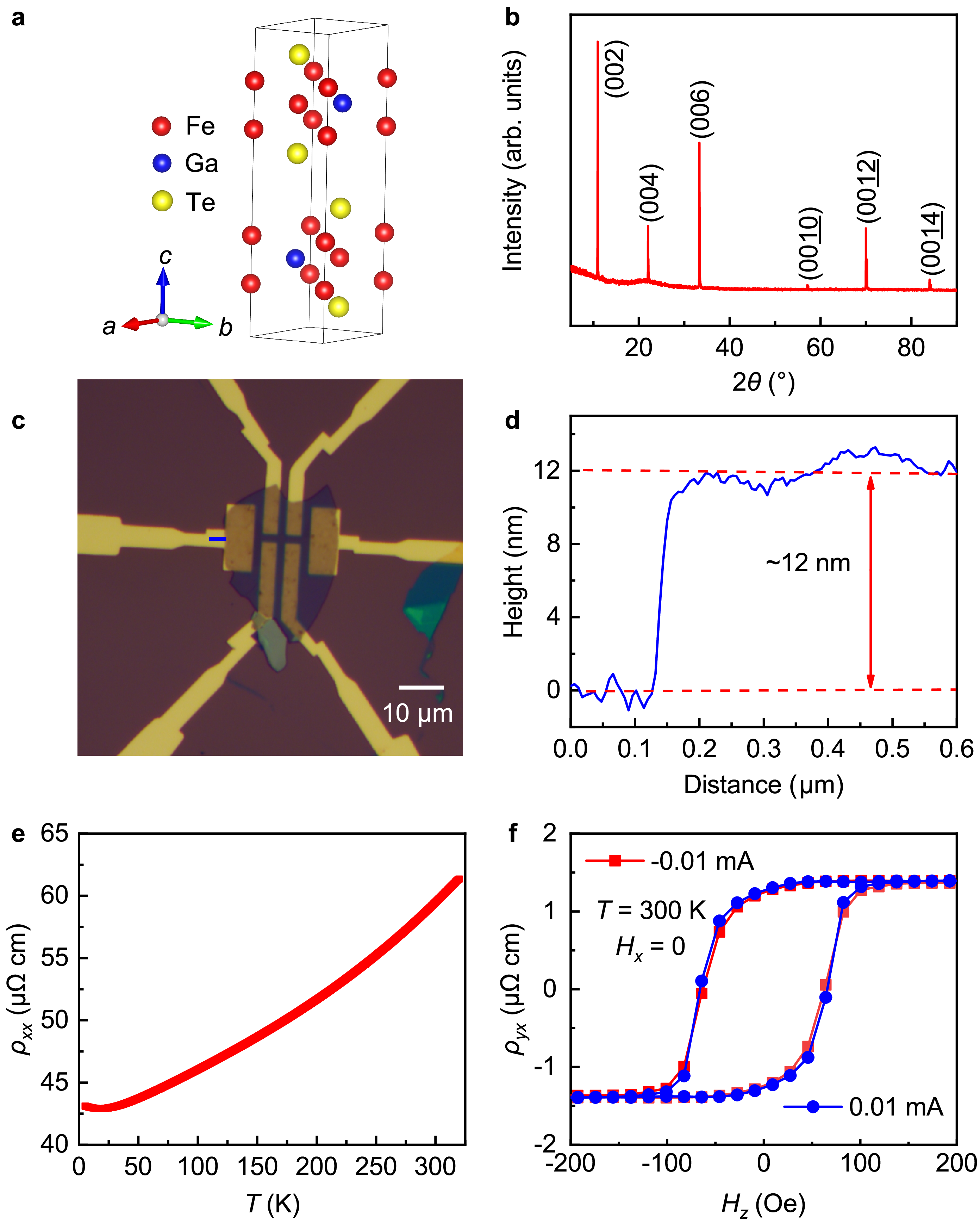}}
\caption{\textbf{Crystal structure and transport properties of a typical Fe$_{3}$GaTe$_{2}$  thin flake.} 
\textbf{a}, Crystal structure of Fe$_{3}$GaTe$_{2}$. The red, blue and yellow balls represent Fe, Ga and Te atoms, respectively.
\textbf{b}, XRD pattern of  a Fe$_{3}$GaTe$_{2}$ single crystal.
\textbf{c}, Optical image of a Fe$_{3}$GaTe$_{2}$ thin flake on a 285 nm SiO$_{2}$/Si substrate. The white scale bar represents 10 $\mu$m. 
\textbf{d}, Cross-sectional profile of the Fe$_{3}$GaTe$_{2}$ thin flake along the blue line in c.
\textbf{e}, Temperature dependence of $\rho_{xx}(T)$ for Fe$_{3}$GaTe$_{2}$ with $t$ = 12 nm.
\textbf{f}, Field dependence of $\rho_{yx}(H_{z})$ at 300 K with $ab$-plane currents of  +0.01 mA and -0.01 mA .}
\label{fig:boat1}
\end{figure}

\begin{figure}
\centerline{\includegraphics[width=0.8\linewidth]{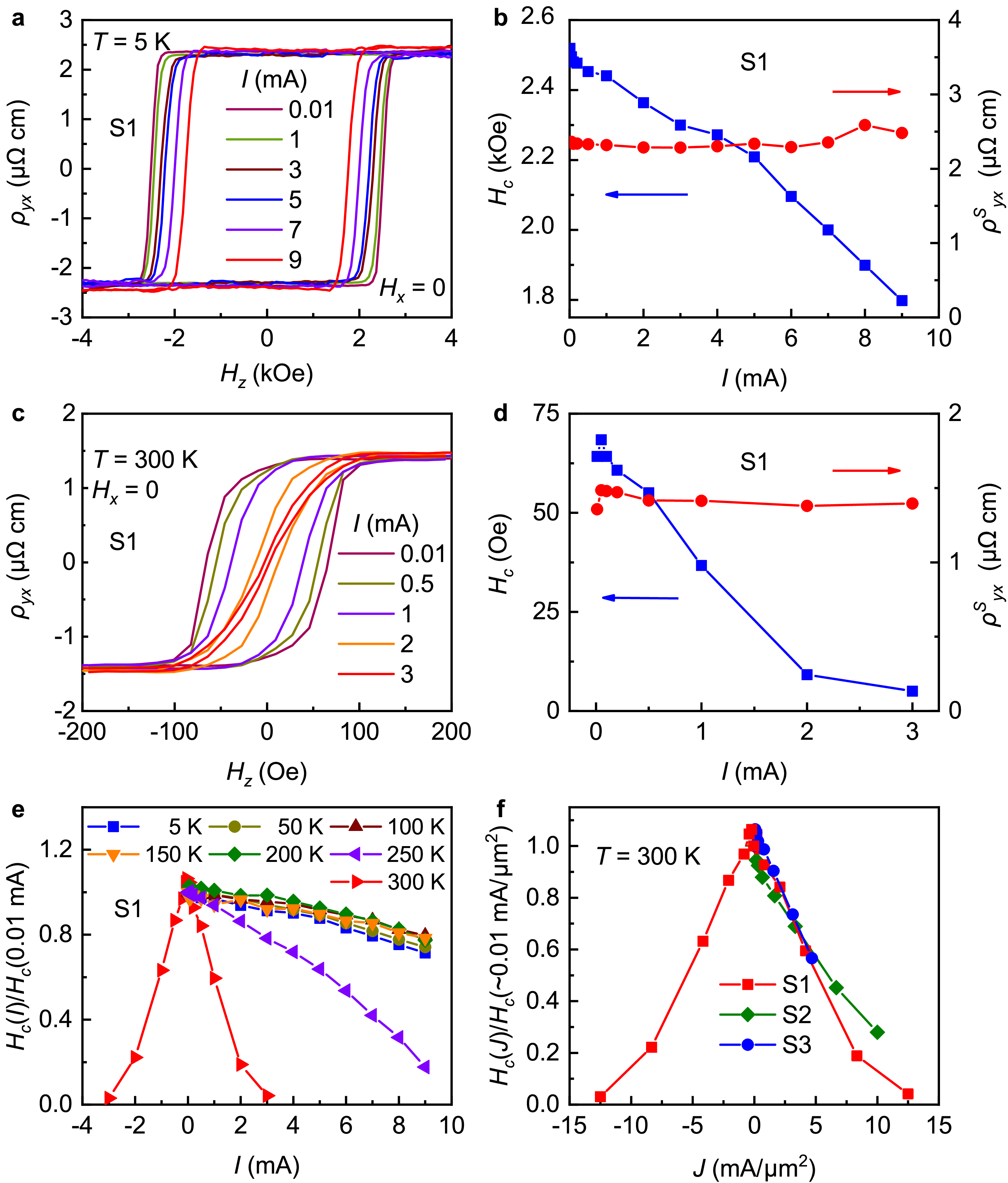}}
\caption{\textbf{Evolution of $\rho_{yx}(H_{z})$ hysteresis loop with $I$ in Fe$_{3}$GaTe$_{2}$ thin flakes.} 
\textbf{a}, Field dependence of $\rho_{yx}(H_{z})$ with an $ab$-plane $I$ from 0.01 mA to 9 mA for sample 1 (S1) at 5 K .
\textbf{b}, the $H_{c}$ and $\rho_{yx}^{s}$ as a function of $I$ for S1 at 5 K.
\textbf{c}, Field dependence of $\rho_{yx}(H_{z})$ at various $I$ at 300 K. 
\textbf{d}, the $I$ dependence of $H_{c}$ and $\rho_{yx}^{s}$ for S1 at 300 K.
\textbf{e}, the $I$ dependence of $H_{c}(I)/H_{c}(I = 0.01 \rm mA)$ at different temperatures for S1.
\textbf{f}, the $H_{c}(J)/H_{c}(J = 0.01 \rm mA/\mu m^{2})$ as a function of $J$ for three devices at 300 K.
For all of these measurements, the $H_{x}$ = 0 Oe.}

\label{fig:boat1}
\end{figure}

\begin{figure}
\centerline{\includegraphics[width=0.8\linewidth]{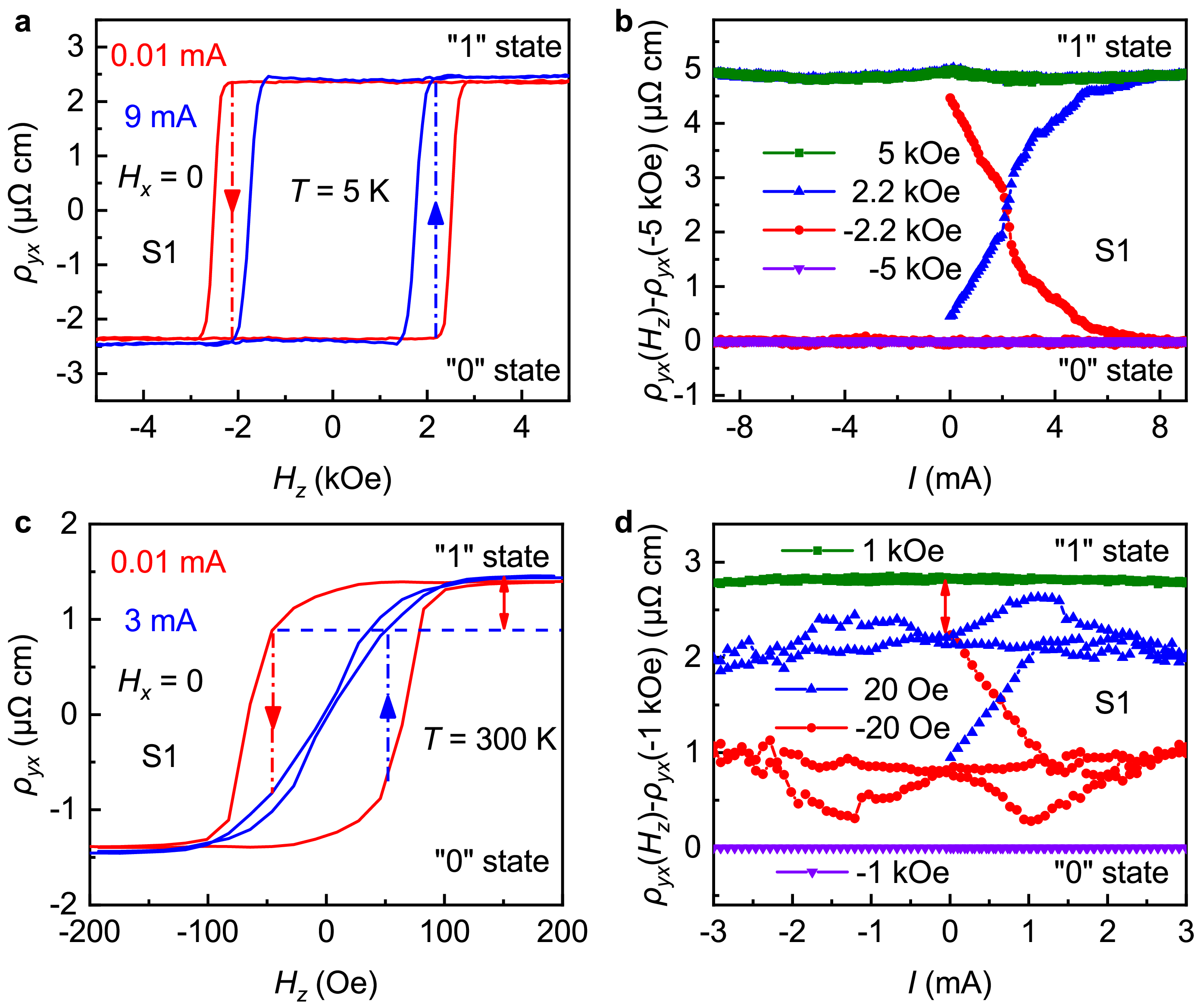}}
\caption{\textbf{Nonvolatile magnetization switching by current.} 
\textbf{a}, $\rho_{yx}(H_{z})$ curves of sample S1 under $I$ = 0.01 mA and 9 mA at $T$ = 5 K. The blue and red arrows indicate the transitions between ``0" and ``1" states driven by current at $H_{z}$ = $\pm$2.2 kOe.
\textbf{b},  $\rho_{yx}(H_{z})$ - $\rho_{yx}$(-5 kOe) as a function of current $I$  path (0 $\rightarrow$ 9 $\rightarrow$ -9 $\rightarrow$ 9 mA) under various fields from -5 kOe to 5 kOe at 5 K. The red and blue curve shows the switching from ``1" to ``0" and ``0'' to ``1'', respectively.
\textbf{c}, $\rho_{yx}(H_{z})$ curves of sample S1 under $I$ = 0.01 mA and 3 mA at $T$ = 300 K .
\textbf{d}, $\rho_{yx}(H_{z})$ - $\rho_{yx}$(-1 kOe) as a function of current $I$  path (0 $\rightarrow$ 3 $\rightarrow$ -3 $\rightarrow$ 3 mA) under various fields from -1 kOe to 1 kOe at 300 K. The red and blue curve shows the switching from ``1" to ``0" and ``0'' to ``1'', respectively. For all of these measurements, the $H_{x}$ = 0 Oe.}
\label{fig:boat1}
\end{figure}

\begin{figure}
\centerline{\includegraphics[width=0.8\linewidth]{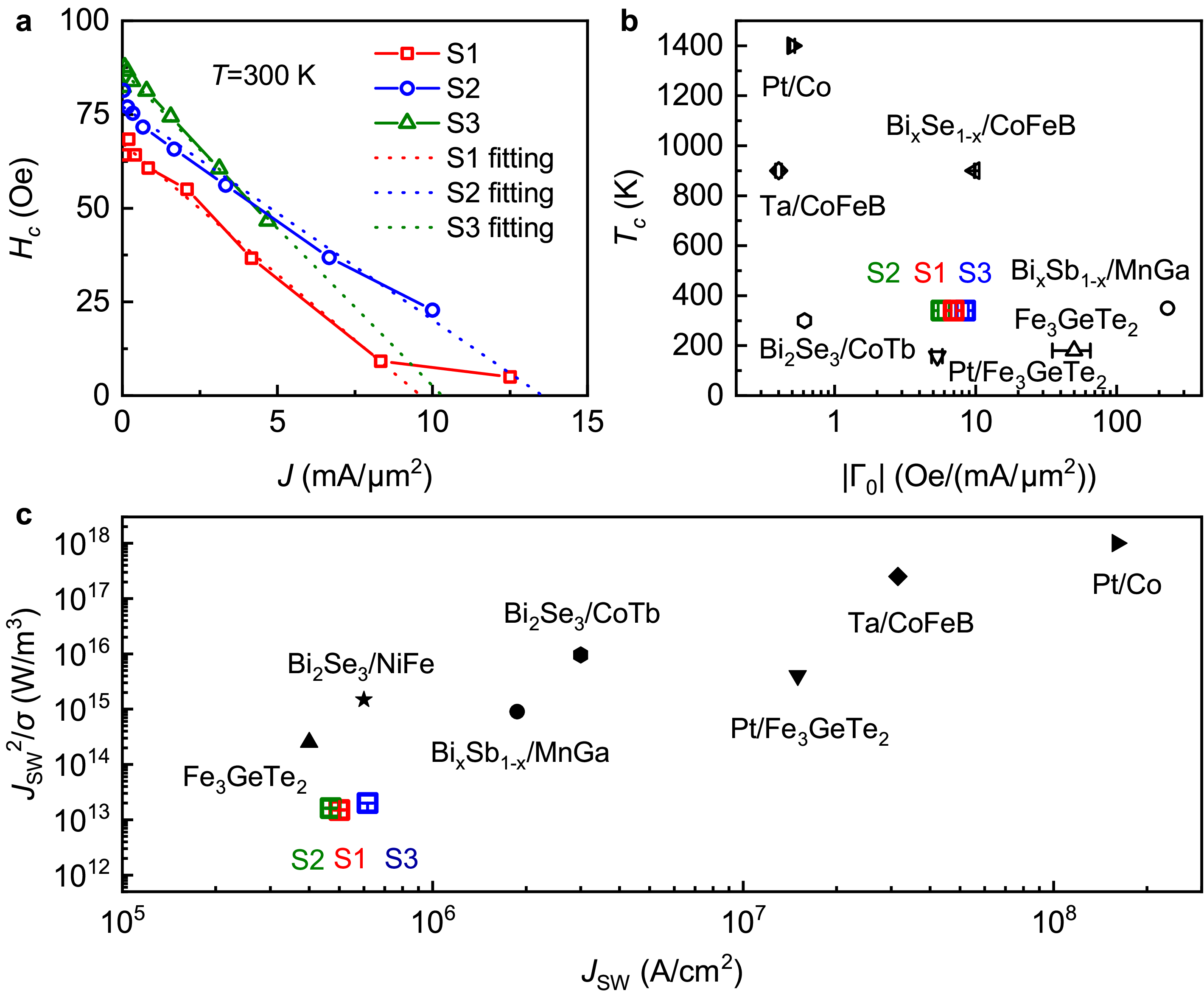}}
\caption{\textbf{Highly efficient magnetization switching by current at 300 K.} 
\textbf{a}, $J$ dependence of $H_{c}$ at 300 K and linear fits at low current density range for S1-S3.
\textbf{b}, $|\Gamma_{0}|$ and $T_{c}$ of Fe$_{3}$GaTe$_{2}$ and other SOT systems \cite{ZhangKX,WangX,Khang,HanJ,Garello,M}.
\textbf{c}, Switching current density $J_{\rm SW}$ and switching power dissipation $J_{\rm SW}^{2}/\sigma$ of Fe$_{3}$GaTe$_{2}$ and various SOT systems \cite{ZhangKX,WangX,Khang,HanJ,Garello,M,WangY}.
The data for Fe$_{3}$GeTe$_{2}$ and Pt/Fe$_{3}$GeTe$_{2}$ is measured at 2 K and 140 K, respectively, when other data are obtained at 300 K.}
\label{fig:boat1}
\end{figure}

\newpage
\begin{figure}
	\centerline{\includegraphics[width=0.8\linewidth]{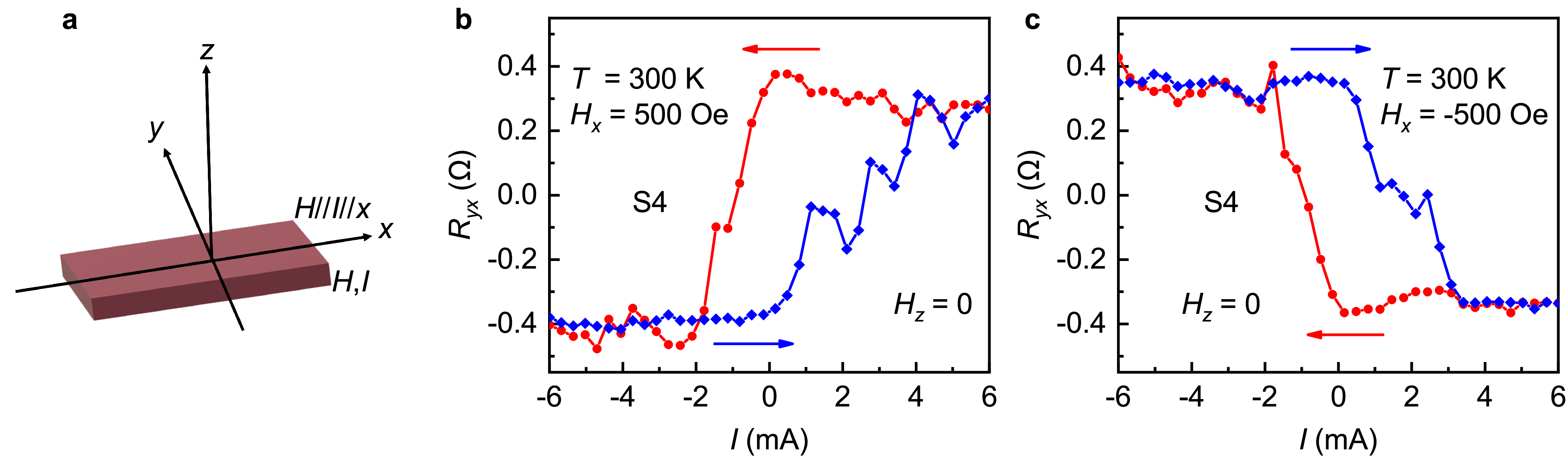}}
	\caption{\textbf{Magnetization switching with an $ab$-plane magnetic field $H_{x}$.} 
	\textbf{a}, The schematic diagram of the Fe$_{3}$GaTe$_{2}$ device. Both current and magnetic field are along the $x$ direction.
	\textbf{b} and \textbf{c},  Current-driven magnetization switching with $H_{x}$ = 500 Oe and -500 Oe at 300 K and $H_{z}$ = 0 Oe. The switching polarity is anticlockwise and clockwise, respectively. }
	\label{fig:boat1}
\end{figure}

\end{document}